\documentstyle[twocolumn,prl,aps]{revtex}

\input epsf

\begin{document}
\title{Resonant magnetic field control of elastic scattering in cold $^{85}$Rb}
\author{J. L. Roberts, N. R. Claussen, James P. Burke, Jr., Chris H. Greene, 
E. A. Cornell,* and C. E. Wieman}
\address{JILA and the Department of Physics, University of Colorado, Boulder, 
CO\\ 80309-0440}
\date{\today}
\maketitle

\begin{abstract}
A magnetic field dependent Feshbach resonance has been observed in the
elastic scattering collision rate between atoms in the $F=2,M=-2$ state of $%
^{85}$Rb. Changing the magnetic field by several Gauss caused the collision
rate to vary by a factor of 10$^{4}$, and the sign of the scattering
length could be reversed. The resonance peak is at 155.2(4) G and its width
is 11.6(5) G. From these results we extract much improved values for the
three quantities that characterize the interaction potential:
the van der Waals coefficient C$_6$, the singlet scattering length $a_{S}$, 
and the triplet scattering length $a_{T}$. 
\end{abstract}


Very low-temperature collision phenomena can be quite sensitive to applied
electromagnetic fields. Several groups have altered inelastic collision
rates in optical traps by applying laser fields\cite{Gould96}. There have
also been numerous proposals \cite
{Tiesinga9293,Marinescu98,Fedichev96,BohnJulienne97,Moerdijk95} 
for using laser and
static electric or magnetic fields to influence the S-wave scattering length 
($a$%
), and equivalently, the elastic collision cross sections ($\sigma =8\pi
a^{2}$) between cold atoms. Particularly notable is the prediction by
Verhaar and co-workers \cite{Tiesinga9293} that as a function of magnetic
field there should be Feshbach resonances in collisions between cold ($\sim
\mu $K) alkali atoms. These resonances were predicted to cause dramatic
changes in the cross section, and to even allow the sign of the
scattering length to be changed. Such resonances are very
interesting collision physics, but they also offer a means to manipulate
Bose-Einstein condensates (BEC), the properties of which are primarily
determined by the scattering length. Some time ago we searched for such 
resonances in $^{133}$Cs and 
$^{87}$Rb, without success\cite{Monroe93,Newbury95}. This was not surprising
because there was enormous uncertainty as to the positions and widths of the
predicted resonances. However, recent photoassociation spectroscopy has 
greatly improved the knowledge of the alkali interatomic
interaction potentials, allowing these resonance parameters to be predicted
with much less uncertainty \cite{Julienne96}. As we show below, the fact that
the widths and positions of these resonances are so sensitive to the
interatomic potentials makes their measurement a good way to determine these 
potentials.

The improved predictions for the resonance positions facilitated their
observation. In the past few months Feshbach resonances have been
observed in both $^{23}$Na \cite{ketterle} and $^{85}$Rb \cite{heinzen}. In
the sodium work, two magnetic field dependent processes were observed: 1) a
change in the expansion rate of BEC due to a change in
the scattering length, and 2) an enormous, and as yet unexplained, increase
in the loss rate. This loss rate precluded the study
of collisions in the interesting field regime near the center of the
resonance where the scattering length changes sign; the width of the
resonance was not measured. In the rubidium work \cite{heinzen} the resonance 
was detected as a magnetic field dependence of the photoassociation spectra. 
The resulting resonance width was measured to be far larger than originally
predicted.

Here we report the study of a Feshbach resonance in the elastic collision
cross section between atoms in the $F=2,M=-2$ state of $^{85}$Rb. By
changing the DC magnetic field we are able to change the collision rate by
4 orders of magnitude, and explore regions of
positive, negative, and essentially zero scattering length. We
determine the width and position of the resonance about a factor of ten more
precisely than in Ref.\cite{heinzen}, and from this data we improve the 
accuracy of the Rb interaction potential substantially. In contrast to 
Ref.\cite{ketterle}, 
we do not observe two or three body loss at resonance because we work at 
much lower densities ($10^{9}$ atoms/cm$^{3}$).

We measured the collision rates using the technique of ``cross dimensional
mixing,'' as in our earlier work\cite{Monroe93}. In this technique, a
nonisotropic distribution of energy is created in a magnetically trapped
cloud of atoms, and the time for the cloud to reequilibrate by elastic
collisions is measured. The apparatus used is identical to that used in our
previous work on BEC in $^{87}$Rb\cite{Myatt96}. It is a double magneto-optic 
trap (MOT) system in which multiple samples of atoms are trapped in a 
relatively high-pressure
chamber, and then transferred to a second MOT in a low-pressure chamber.  
The second MOT is then turned off and a ``baseball coil'' magnetic trap is
turned on around them. The atoms are then cooled by forced evaporation.

After the atoms are evaporatively cooled to the desired
temperature, the energy in the radial direction (and correspondingly the
square of the width of the trapped cloud) is reduced to about 0.6 that of
the axial direction. This is done by decreasing the frequency
of the RF ``knife'' used for the evaporative cooling more rapidly than the 
cloud can equilibrate. The bias magnetic field is then adiabatically ramped 
to the selected value,
and the cloud is allowed to equilibrate for a fixed
time. The shape of the cloud is then measured by absorptive imaging. This is
repeated for different equilibration times, and the aspect ratios of the
cloud vs. time are fit to an exponential to determine the equilibration time
constant $\tau _{eq}$. As discussed in Ref.\cite{Monroe92}, $\tau _{eq}$ is
proportional to the inverse of the elastic collision rate\cite{PropConst98}.
From the measured trap oscillation frequencies and the
measured shape and optical depth, we determine the temperature $T$ and 
average density $\langle n \rangle$ of the cloud. The magnetic field value 
at the center of the cloud is
determined to $\pm $0.1 G by finding the RF frequency that resonantly drives
spin flip transitions for the atoms at that position and using the
Breit-Rabi equation. The spread in magnetic field across the cloud scales 
as $T^{1/2}$, and is 0.6 G FWHM for a 0.5 $\mu $K cloud.

The measured equilibration times depend on $B$ and $T$ and vary from 0.15 s 
to nearly 2000 s.  
To display data taken at several values of both density and temperature, we 
convert it to the normalized equilibration rate, $\Gamma_{norm}=1/ \langle nv 
\rangle \tau_{eq}$.  
Figure 1 displays the data versus mean magnetic field.  $\Gamma_{norm}$ has 
units of cross-section and in fact is proportional to an empirical average 
over the field- and energy-dependent 
elastic scattering cross-section $\sigma(B,E)$.  To calculate 
this average is problematic --- a detailed Monte Carlo simulation of particle
trajectories \cite{dalibard_cesium} is probably required. However in
the low-temperature limit it is clear that the maximum and minimum values
of $\Gamma_{norm}$ will occur at the same magnetic field as the peak and
zero of $\sigma(B,E=0)$.    For a rough qualitative comparison, in Fig. 1
we have also plotted the theoretically predicted values of $\sigma(B,E)$ 
for two temperatures. These predictions are calculated using the atomic 
potentials determined from the observed maximum and minimum values of the 
relaxation rate, as discussed below.  Both the cross section curves and the 
equilibration data show the same qualitative features. 
For temperatures below a few $\mu $K, there is a slightly
asymmetric peak near 155 G, and the width and height of this peak are
strong functions of temperature. At ~167 G there is a profound
drop in the rate. This dip is quite asymmetric, but the shape is relatively
insensitive to $T$, and at the bottom (field value $B_{{\rm min}}$) the rate
is essentially zero. The field value of the peak ($B_{{\rm peak}}$) is
customarily defined to be the position of the Feshbach resonance, and the
resonance width, $\Delta$ is the distance between $B_{{\rm peak}}$ and 
$B_{{\rm min}}$\cite{heinzen}.
The scattering length is positive for field values between $B_{{\rm peak}}$ 
and $B_{{\rm min}}$, and is negative for field values below $B_{{\rm peak}
}$ or above $B_{{\rm min}}$. This dependence of the sign is expected from 
previous theory.  The observation that $B_{{\rm peak}}$ is at a lower
field than $B_{{\rm min}}$ provides experimental confirmation that the 
scattering length is negative away from the resonance.

Because the functional form of the normalized equilibration rate is not 
known, the desired
quantities, $B_{{\rm peak}}$ and $B_{{\rm min}}$ cannot be found by a
detailed fit to all of the data. We determine them by fitting a simple smooth
curve to only the few highest (or lowest) points at each temperature below 
5 $\mu $K, and assigning correspondingly conservative error bars that more 
than span the values determined for all temperatures\cite{values98}. 
We find $B_{{\rm peak}}$ = 155.2 (4) G and $B_{\rm min}$ = 166.8 (3) G, giving 
a resonance width $\Delta$ = 11.6 (5) G.  The values for $B_{{\rm peak}}$ 
and $\Delta$ are reasonably consistent with the less precise values 
164(7) and 8.4 (3.7) G measured in Ref.\cite{heinzen}. In our experiment, 
the accuracy of the peak
position is much higher primarily because of better field calibration. The
better accuracy in the width is largely because the photoassociation
measurement is a somewhat less direct way to observe the resonance, and so
substantial and somewhat uncertain corrections are required to go from
observed signal widths to the actual resonance width.

We can now use these measured quantities to determine the singlet and
triplet Born-Oppenheimer potentials. The accuracy of predicted cold
collision observables hinges on the quality of these potentials used in the
radial Schr\"{o}dinger equation for the nuclear motion. Accurate 
determination of these
potentials in turn relies primarily on three parameters, the long range 
van der Waals coefficient C$_{6}$ and the
zero-energy singlet and triplet phases. The phases are usually tabulated in
terms of singlet $a_{S}$ and triplet $a_{T}$ scattering lengths calculated
with hyperfine terms omitted from the radial equation.

The Feshbach resonance involves states with both singlet and triplet
character; consequently the present determination of resonance
width and position versus $B$ can be translated with the aid of theory into
greatly improved values for the singlet and triplet scattering lengths. The
old ``nominal'' values of the scattering lengths and C$_{6}$ (optimized to
achieve agreement with previous measurements) predicted the position of this
Feshbach resonance reasonably well, but, as in Ref. \cite{heinzen} the 
predicted width was much smaller
than observed. This width reflects the coupling of bound and continuum
channels, and is primarily controlled by the difference between the singlet
and triplet scattering lengths. It is inconvenient to tabulate
results as a function of singlet and triplet scattering lengths for $^{85}$
Rb, because both are very close to lying on top of a divergent pole.
Accordingly, we present the results of our analysis in terms of a
dissociation phase $\nu_{{\rm D}}$ as in Ref.\cite{heinzen}. 
We define the corresponding singlet $\nu_{{\rm DS}}$ or triplet 
$\nu_{{\rm DT}}$ dissociation phase
in terms of the scattering length through the relation\cite{Mies96}: 
$\nu_{{\rm D}} = (1 / \pi) \cot^{-1}
(a / a_{{\rm ref}} - 1)$ where 
$a_{{\rm ref}}=\Gamma(5/4)/ [\sqrt{8}\,\Gamma(3/4)]
(2m {\rm C_{6}}/\hbar^{2})^{1/4}$ \cite{Flaumbaum}
and $m$ is the
reduced mass of the atomic pair. The dissociation phase is related to the 
short range quantum defect $\mu^{{\rm sr}}$ 
presented in Ref.\cite{QDT98} by $\nu_{{\rm D}} \cong \mu^{{\rm sr}} - 1/4$. 
The quantum defect method developed in Ref.\cite{QDT98} was used to 
generate the theory curves in Fig. 1.

We have adjusted the singlet and triplet Rb-Rb potential curves at small 
distances until the theoretical 
calculation agrees simultaneously with the present measurements of 
$B_{{\rm peak}}$ and $\Delta$, to within
the experimental uncertainties.  For a given C$_{6}$, this severely 
constrains the values of both
$\nu_{{\rm DS}}$ and $\nu_{{\rm DT}}$ (see Fig. 2).  However, the position 
and width are not sufficient 
to totally constrain all three parameters C$_{6}$, $\nu _{{\rm DS}}$, and 
$\nu _{{\rm DT}}$.
In order to obtain values for all the parameters, we also require that our 
triplet potentials predict the $g$-wave shape resonance in an energy range 
consistent with the measured value given in Ref.\cite{Boesten96}.

The results of this analysis are presented in Fig. 2.
It is immediately clear from the figure that accurate measurement
of Feshbach resonances are an extremely precise method for determining
threshold properties of the inter-atomic potentials. In fact, our
measurement has allowed us to reduce the combined C$_{6}$, 
$\nu_{{\rm DS}}$, and $\nu_{{\rm DT}}$ parameter space by roughly a factor 
of 80.
As discussed in Ref.\cite{heinzen}, the position of the resonance
depends mostly on C$_{6}$ and the sum $\nu_{{\rm DS}} + \nu_{{\rm DT}}$. 
The width depends mostly on the difference $\nu_{{\rm DS}} - \nu_{{\rm DT}}$.
As Fig. 2 shows, the allowed $\nu_{{\rm DS}}$, $\nu_{{\rm DT}}$ parameter
region is an extremely correlated function of C$_{6}$. In particular, we
find that the nominal value of the
sum and difference coordinates are linear functions of C$_{6}$
and the uncertainties in the coordinates are constant. 
Accordingly, we can represent the allowed parameter region
in the following manner:\\
$\nu_{{\rm DS}} + \nu_{{\rm DT}}$ = -0.0442 + 2.2(10$^{-4}$)(C$_{6}$ - 
\={C}$_{6})\pm 0.0005$ and
$\nu_{{\rm DS}} - \nu_{{\rm DT}}$ = 0.0666 - 6.2(10$^{-5}$)(C$_{6}$ - 
\={C}$_{6})\pm 0.003$ where \={C}$_{6}$ = 4700.
Converting the dissociation phases into scattering lengths, we find reasonable
agreement with previous work\cite{Burke98,Tsai97,Burke97}.  We find for the 
nominal C$_{6}$ the values (in a.u.)
a$_t$($^{85}$Rb)=$-$363$\pm$10, a$_s$($^{85}$Rb)=2300$^{+300}_{-200}$, 
a$_t$($^{87}$Rb)=109.3$\pm$0.4, and a$_s$($^{87}$Rb)=92.7$\pm$0.4. 
Our value for C$_{6}$, 4700(50) a.u. is slightly higher (and with smaller 
uncertainty) than a previous analysis based solely on the $g$-wave shape 
resonance \cite{Boesten96}.
We have confirmed that several scattering observables predicted by the 
new Born-Oppenheimer potentials are
consistent with previous measurements.  Specifically, the new potentials 
predict a broad $d$-wave shape resonance\cite{Boesten97} in $^{87}$Rb, 
the scattering length ratio $a_{2,1}/a_{1,-1}$\cite{Hall98} and the 
thermally-averaged $|2,2\rangle $+$|1,-1\rangle $ inelastic rate 
constant\cite{Myatt97} that are consistent with previous measurements.
We also find 10 of the 12 measured $d$-wave bound states\cite{Tsai97} 
within the 2$\sigma $ error bars\cite{Heinzen_Private}. The new
potentials also permit us to predict additional $^{85}$Rb Feshbach
resonances\cite{Moerdijk95}, at $B_{{\rm peak}}$= 226.5 $\pm $ 4.8 G 
with a width of ~0.01G, and at $B_{{\rm peak}}$= 535.8 $\pm $ 4.0 G 
with a width of 2.2 $\pm $0.2 G.

We have shown that precise measurements of the Feshbach resonance in the
elastic scattering cross section provide unprecedented knowledge of the
Born-Oppenheimer potentials that control Rb-Rb scattering processes at
sub-mK temperatures. We have also used this resonance to enhance evaporative
cooling of $^{85}$Rb in a low density magnetic trap, and change the
scattering length from negative to positive. These capabilities should allow
novel studies of BEC in the future.

We are pleased to acknowledge valuable conversations with Dan Heinzen,
Boudwijn Verhaar, and John Bohn, as well as the assistance of Richard Ghrist
and Eric Burt in the early stages of the work. NSF, ONR, and NIST supported
this work.

*Quantum Physics Division, National Institute of Standards and Technology

\begin{figure}
	{\caption {(a) The data points show the measured equilibration 
rate divided by average density and velocity 
($\Gamma_{norm} = 1/ \langle nv \rangle \tau_{eq}$) 
vs. magnetic field $B$.  Data are shown for clouds	
at five different temperatures:  0.3 $\mu$K (filled triangles), 0.5 $\mu$K
($\bullet$), 1.0 $\mu$K ($\circ$), 3.5 $\mu$K ($\bigtriangledown$), and 
9.0 $\mu$K (filled squares).  The lines are calculations for the 
thermally averaged elastic cross sections ({\bf not} equilibration rates!) 
and hence are not expected to fit the data points.  The solid line
corresponds to 0.5 $\mu$K and the dot-dashed to 9.0 $\mu$K.  (b),(c)  
Expanded views of the regions of maximum and minimum cross section.}}
\end{figure}

\begin{figure}
\par
\caption {Comparison of the allowed $\nu_{{\rm DS}}$, $\nu_{{\rm DT}}$
parameter space based on this work and previous measurements for a 
constant value of C$_{6}$.  The rectangle is the allowed range
from Ref. [22].  The large diamond is the range from
Ref. [11].  The small filled diamond is the range from this work.
Note that the rectangle and large diamond constraints were established
using a different value of C$_{6}$ from the one used in this work (4550 
instead of 4700 a.u.).  As is discussed in the text, the position of the 
constrained regions depends linearly upon C$_{6}$, which is indicated by
the small connected diamonds on either side of the small filled diamond.
These diamonds show the effect of 50 a.u. uncertainty in C$_{6}$.  The 
arrows indicate sensitivity to changes in the resonance width and position
in this paramter space.}
\end{figure}


\noindent

\begin{references}

\bibitem{Gould96} V. Sanchez-Villicana, S. D. Gensemer, and P. L. Gould,
Phys. Rev. A {\bf 54}, R3730 (1996); and references therein.

\bibitem{Tiesinga9293}  E. Tiesinga, A. J. Moerdijk, B. J. Verhaar, and H.
Stoof, Phys. Rev. A {\bf 46}, R1167 (1992); E. Tiesinga, B. J. Verhaar, and
H. Stoof, Phys. Rev. A {\bf 47}, 4114 (1993).

\bibitem{Marinescu98}  M. Marinescu and L. You (submitted to PRL).

\bibitem{Fedichev96}  P. O. Fedichev {\it et al.}, Phys. Rev. Lett. 
{\bf 77}, 2913 (1996).

\bibitem{BohnJulienne97}  J. L. Bohn and P. S. Julienne, Phys. Rev. A 
{\bf 56}, 1486 (1997).

\bibitem{Moerdijk95}  A. J. Moerdijk {\it et al.}, Phys. Rev. A {\bf 51},
4852 (1995); J. M. Vogels {\it et al.}, Phys. Rev. A {\bf 56}, R1067 (1997).

\bibitem{Monroe93}  C. R. Monroe, E. A. Cornell, C. A. Sackett, C. J. Myatt,
and C. E. Wieman, Phys. Rev. Lett. {\bf 70}, 414 (1993).

\bibitem{Newbury95}  N. R. Newbury, C. J. Myatt, and C. E. Wieman, Phys.
Rev. A {\bf 51}, R2680 (1995).

\bibitem{Julienne96}  For a recent review on this subject, see for instance:
P. S. Julienne, J. Res. Natl. Inst. Stand. Technol. {\bf 101}, 487 (1996).

\bibitem{ketterle}  {S. Inouye, M. R. Andrews, J. Stenger, H. J. Miesner, D.
M. Stamper-Kurn, and W. Ketterle, Nature {\bf 392}, 151 (1998).}

\bibitem{heinzen}  {Ph. Courteille, R. S. Freeland, and D. J. Heinzen, 
F. A. van Abeelen, and B. J. Verhaar, 
Phys. Rev. Lett. {\bf 81}, 69 (1998).}

\bibitem{Myatt96}  C. J. Myatt, N. R. Newbury, R. W. Ghrist, S.
Loutzenhiser, and C. E. Wieman, Opt. Lett. {\bf 21}, 290 (1996).

\bibitem{Monroe92}  C. R. Monroe, 1992 Doctoral Dissertation in Physics,
University of Colorado.

\bibitem{PropConst98}  The proportionality constant is about
2.6 when the elastic cross section is temperature-independent, which
is decidedly not the case here.

\bibitem{dalibard_cesium}  D. Gury-Odelin, J. Sding, P. Desbiolles, and 
J. Dalibard, Optics Express {\bf 2}, 323 (1998).

\bibitem{values98}  The three values of $B_{{\rm peak}}$ determined for the
0.5, 1, and 0.75 $\mu $K data (the latter not shown in Fig. 1 to avoid 
cluttering it excessively) can be determined to about the same precision 
and agree to within 0.4 G.
The value of $B_{{\rm min}}$ is primarily based on the 3 $\mu $K data. 

\bibitem{Mies96} F. H. Mies, C. J. Williams, P. S. Julienne, and M. Krauss,
J. Res. Natl. Inst. Stand. Technol. {\bf 101}, 521 (1996).

\bibitem{Flaumbaum} G. F. Gribakin and V. V. Flambaum, Phys. Rev. A
{\bf 48}, 546 (1993).

\bibitem{QDT98}  J. P. Burke, Jr., C. H. Greene, and J. L. Bohn, (submitted
to PRL, June 1998). The scattering length is given in terms
of the quantum defect by: $a={-C^{2}\tan (\pi \mu^{\rm sr} )}%
/[1+{\cal G}(0)\tan (\pi \mu^{\rm sr} )]$, where
$C^{2}=0.957217(2m{\rm C}_{6}/\hbar^{2})^{1/4}$ and ${\cal G}(0)=-1.00260$.

\bibitem{Boesten96}  H.M.J.M. Boesten, C. C. Tsai, B. J. Verhaar, and D. J.
Heinzen, Phys. Rev. Lett. {\bf 77}, 5194 (1996).

\bibitem{Burke98} J. P. Burke, Jr., J. L. Bohn, B. D. Esry, and C. H. Greene,
Phys. Rev. Lett. {\bf 80}, 2097 (1998). 
\bibitem{Tsai97}  C. C. Tsai, R. S. Freeland, J. M. Vogels,
H.M.J.M. Boesten, B. J. Verhaar, and D. J. Heinzen, Phys. Rev. Lett. 
{\bf 79}, 1245 (1997).

\bibitem{Burke97}  P. S. Julienne, F. H. Mies, E. Tiesinga, and C. J. 
Williams, Phys. Rev. Lett. {\bf 78}, 1880 (1997);
J. P. Burke, Jr., J. L. Bohn, B. D. Esry, and C.
H. Greene, Phys. Rev. A {\bf 55}, R2511 (1997).

\bibitem{Heinzen_Private} Those in disagreement are likely due to somewhat
overly optimistic error bars for the positions of the bound states
as determined by photoassociation experiments. (Dan Heinzen, 
private communication).

\bibitem{Boesten97} H.M.J.M. Boesten, C. C. Tsai, J. R. Gardner, D.
J. Heinzen, and B. J. Verhaar, Phys. Rev. A {\bf 55} 636 (1997). 

\bibitem{Hall98} M. R. Matthews, D. S. Hall, D. S. Jin, J. R. Ensher,
C. E. Wieman, E. A. Cornell, F. Dalfovo, C. Minniti, and S. Stringari,
Phys. Rev. Lett. {\bf 81}, 243 (1998).

\bibitem{Myatt97}  C. J. Myatt, E. A. Burt, R. W. Ghrist, E. A.
Cornell, and C. E. Wieman, Phys. Rev. Lett. {\bf 78}, 586 (1997).
\end{references}
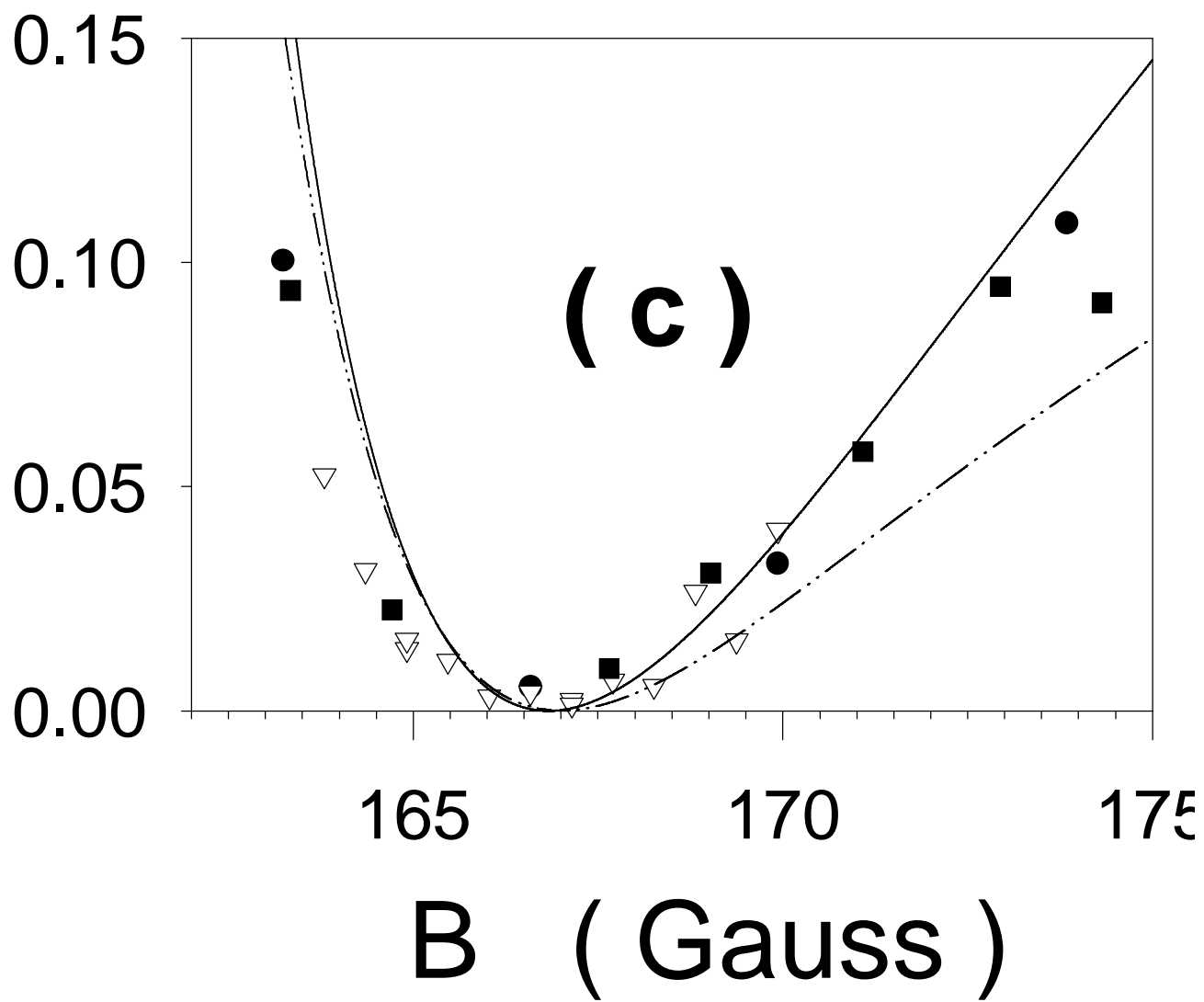
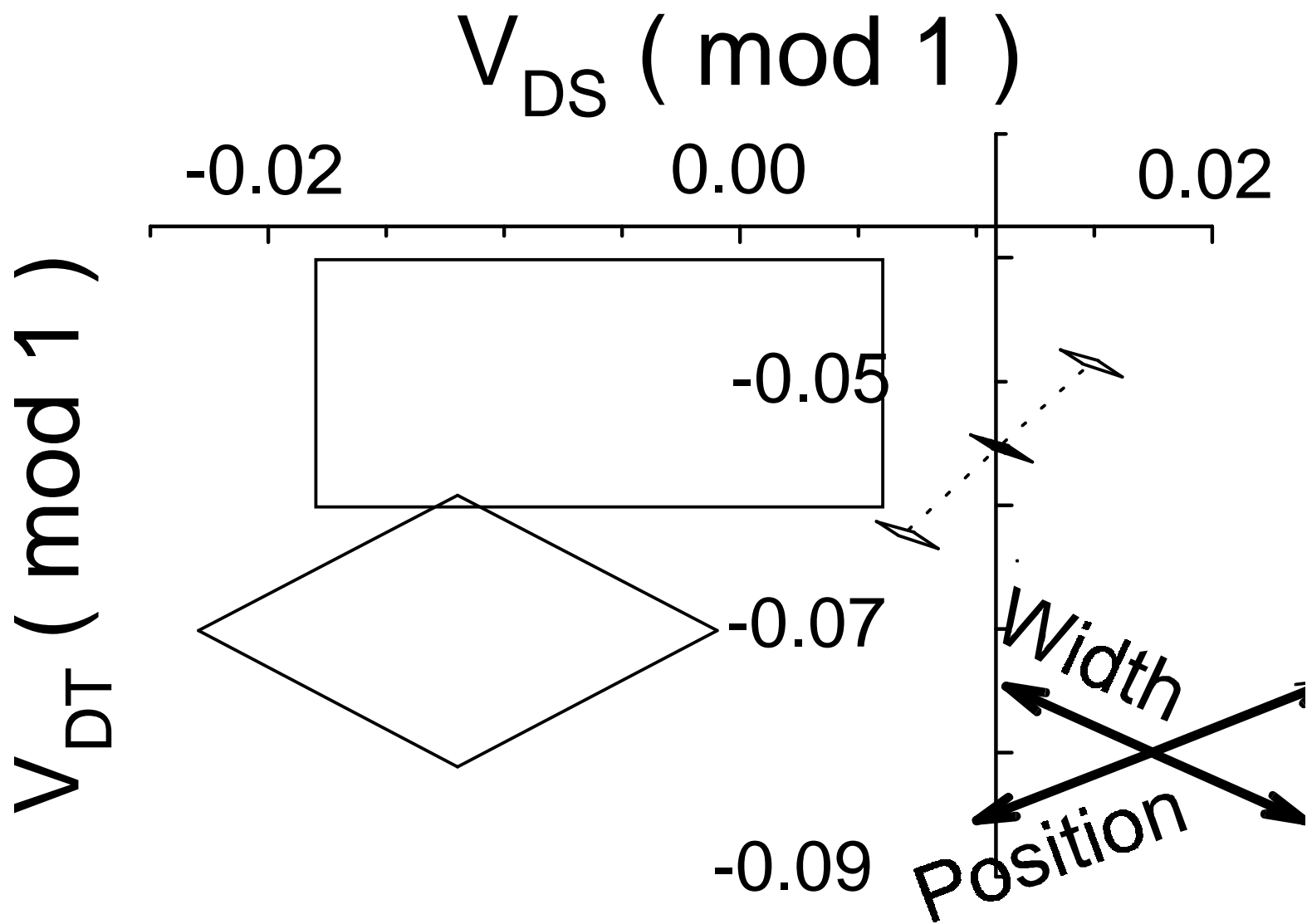
\end{document}